\documentstyle[12pt,procsla]{article}
\begin{document}

\thispagestyle{empty}
\vspace*{-1.9cm}
\begin{flushright}
{\small TTP96-25\\
hep-ph/9606469
}
\end{flushright}

\begin{center}
{\bf CP VIOLATION IN THE B SYSTEM\footnote{Invited talk given at
the {\it III German--Russian Workshop on Heavy Quark Physics}, Dubna, 
Russia, May 20--22, 1996, to appear in the proceedings.}
\\}

\vspace*{1cm}
ROBERT FLEISCHER\\
{\it Institut f\"ur Theoretische Teilchenphysik, Universit\"at
Karlsruhe\\
D--76128 Karlsruhe, Germany\\}
\end{center}

\vspace*{0.85cm}
\begin{abstracts}
{\small  Strategies for extracting CKM phases from non-leptonic $B$ decays
are reviewed briefly. Both general aspects and some recent developments
including CP-violating asymmetries, the $B_s$ system in light of a
possible width difference $\Delta\Gamma_s$, and triangle relations among
$B$ decay amplitudes are discussed. The role of electroweak penguins
in these strategies is illustrated briefly.}
\end{abstracts}

\vspace*{0.25cm}
\section{Setting the Scene}\label{intro}
Within the Standard Model of electroweak interactions\cite{sm}, CP violation
is closely related to the Cabibbo--Kobayashi--Maskawa matrix\cite{cab,km}
(CKM matrix) connecting the electroweak eigenstates of the $d$-, $s$- and
$b$-quarks with their corresponding mass eigenstates: 
\begin{equation}\label{e1}
\left(\begin{array}{c}
d\\
s\\
b
\end{array}\right)_{\mbox{{\scriptsize EW}}}=\left(\begin{array}{ccc}
V_{ud}&V_{us}&V_{ub}\\
V_{cd}&V_{cs}&V_{cb}\\
V_{td}&V_{ts}&V_{tb}
\end{array}\right)\cdot
\left(\begin{array}{c}
d\\
s\\
b
\end{array}\right)_{\mbox{{\scriptsize mass}}}.
\end{equation}
Whereas a single real parameter -- the Cabibbo angle 
$\Theta_{\mbox{{\scriptsize C}}}$ -- suffices to 
parametrize the CKM matrix in the case of two fermion
generations\cite{cab}, three generalized Cabibbo-type angles and a single
{\it complex phase} are needed in the three generation case\cite{km}. This
complex phase is the origin of CP violation within the Standard Model. 

The $B$-meson system is expected to provide a very fertile ground for
testing the Standard Model description of CP violation\cite{cp-revs}. 
Concerning such tests, the central target is the ususal ``non-squashed'' 
unitarity triangle\cite{ut} of the CKM matrix which is a graphical 
illustration of the fact that the CKM matrix is unitary.
At present the unitarity triangle can only be constrained 
indirectly\cite{bloal,bbl-rev} through experimental data from 
CP-violating effects in the neutral $K$ meson system, 
$B^0$--$\overline{B^0}$ mixing and from certain tree 
decays measuring $|V_{cb}|$ and $|V_{ub}|/|V_{cb}|$. 
It should, however, be possible to determine independently the three 
angles $\alpha$, $\beta$ and $\gamma$ of the unitarity triangle directly at 
future $B$ physics facilities\cite{nak}
by measuring CP-violating effects in $B$ decays\cite{cp-revs}. 
Obviously one of the most exciting questions related to these 
measurements is whether the results for $\alpha$, $\beta$, $\gamma$ 
will agree one day or will not. The latter possibility would 
signal ``new'' physics\cite{newphys} beyond the Standard Model.

As far as CP-violating phenomena in the $B$ system and strategies for 
extracting CKM phases are concerned, the key role is played by 
{\it non-leptonic} $B$ decays. In order to analyze such transitions 
theoretically, one uses low energy effective
Hamiltonians that are calculated by making use of the {\it operator
product expansion} yielding transition matrix elements of the
structure
\begin{equation}\label{e2}
\langle f|{\cal H}_{\mbox{{\scriptsize eff}}}|i\rangle=\sum\limits_k C_k(\mu)
\langle f|Q_k(\mu)|i\rangle.
\end{equation}
The operator product expansion allows one to separate the short-distance
contributions to Eq.~(\ref{e2}) from the long-distance contributions 
which are described by perturbative Wilson coefficient functions $C_k(\mu)$
and non-perturbative hadronic matrix elements $\langle f|Q_k(\mu)|
i\rangle$, respectively. As usual, $\mu$ denotes an appropriate 
renormalization scale. 

In the case of $|\Delta B|=1$, $\Delta C=\Delta U=0$ transitions relevant 
for the following discussion we have
\begin{equation}\label{e3}
{\cal H}_{\mbox{{\scriptsize eff}}}={\cal H}_{\mbox{{\scriptsize 
eff}}}(\Delta B=-1)+{\cal H}_{\mbox{{\scriptsize eff}}}(\Delta B=-1)^\dagger
\end{equation}
with
\begin{equation}\label{e4}
{\cal H}_{\mbox{{\scriptsize eff}}}(\Delta B=-1)=\frac{G_{\mbox{{\scriptsize 
F}}}}{\sqrt{2}}\left[\sum\limits_{j=u,c}V_{jq}^\ast V_{jb}\left\{\sum
\limits_{k=1}^2Q_k^{jq}\,C_k(\mu)+\sum\limits_{k=3}^{10}Q_k^{q}\,C_k(\mu)
\right\}\right].
\end{equation}
Here $\mu={\cal O}(m_b)$, $Q_k^{jq}$ are four-quark operators and the label
$q\in\{d,s\}$ corresponds to $b\to d$ and $b\to s$ transitions. 
The index $k$ distinguishes between ``current-current'' 
$(k\in\{1,2\})$, QCD $(k\in\{3,\ldots,6\})$ and electroweak 
$(k\in\{7,\ldots,10\})$ ``penguin'' operators that are related to tree-level, 
QCD and electroweak penguin processes, respectively. The evaluation 
of such low energy effective Hamiltonians beyond the leading
logarithmic approximation has been reviewed recently in Ref.\cite{bbl-rev}
and the reader is referred to that paper for the technicalities. There also
the four-quark operators are given explicitly and numerical values for 
the Wilson coefficient functions can be found.  Beyond the leading 
logarithmic approximation problems arise from renormalization scheme 
dependences which require to include certain quark-level matrix 
elements at $\mu={\cal O}(m_b)$ in order to have a consistent 
calculation\cite{nlo}. 

\section{Strategies for Extracting CKM Phases}\label{stra}
In this section strategies for extracting CKM phases are reviewed briefly.
I will discuss CP-violating asymmetries, the $B_s$ system in light of
a possible width difference $\Delta\Gamma_s$ between the $B_s$ mass 
eigenstates, and triangle relations among certain non-leptonic $B$ decay
amplitudes.

\subsection{CP-violating Asymmetries in $B_d$ Decays}\label{cpviol}
A particular simple and interesting situation arises if we restrict ourselves
to $B_d$ decays into CP self-conjugate final states $|f\rangle$ satisfying
\begin{equation}\label{e5}
({\cal CP})|f\rangle=\pm|f\rangle.
\end{equation}
In that case the corresponding time-dependent CP asymmetry can be expressed
as 
\begin{eqnarray}
\lefteqn{a_{\mbox{{\scriptsize CP}}}(t)\equiv\frac{\Gamma(B^0_d(t)\to f)-
\Gamma(\overline{B^0_d}(t)\to f)}{\Gamma(B^0_d(t)\to f)+
\Gamma(\overline{B^0_d}(t)\to f)}=}\nonumber\\
&&{\cal A}^{\mbox{{\scriptsize dir}}}_{\mbox{{\scriptsize CP}}}(B_d\to f)
\cos(\Delta M_d\,t)+{\cal A}^{\mbox{{\scriptsize 
mix--ind}}}_{\mbox{{\scriptsize CP}}}(B_d\to f)\sin(\Delta M_d\,t),\label{e6}
\end{eqnarray}
where we have separated the {\it direct} CP-violating contributions from
the {\it mixing-induced} CP-violating contributions which are 
characterized by
\begin{equation}\label{e7}
{\cal A}^{\mbox{{\scriptsize dir}}}_{\mbox{{\scriptsize CP}}}(B_d\to f)\equiv
\frac{1-\left|\xi_f^{(d)}\right|^2}{1+\left|\xi_f^{(d)}\right|^2}
\end{equation}
and
\begin{equation}\label{e8}
{\cal A}^{\mbox{{\scriptsize mix--ind}}}_{\mbox{{\scriptsize
CP}}}(B_d\to f)\equiv\frac{2\,\mbox{Im}\,\xi^{(d)}_f}{1+\left|\xi^{(d)}_f
\right|^2},
\end{equation}
respectively. Here direct CP violation refers to CP-violating effects 
arising directly in the corresponding decay amplitudes whereas mixing-induced
CP violation is related to interference effects between 
$B_d^0$--$\overline{B_d^0}$ mixing and decay processes. The time-dependent 
rates $\Gamma(B^0_d(t)\to f)$ and $\Gamma(\overline{B^0_d}(t)\to f)$
in Eq.~(\ref{e6}) describe the time-evolutions due to 
$B_d^0$--$\overline{B_d^0}$ oscillations for initially, i.e.\ at $t=0$, 
present $B^0_d$ and $\overline{B^0_d}$ mesons, respectively, and $\Delta M_d$
denotes the mass difference of the $B_d$ mass eigenstates. In general the
observable 
\begin{equation}\label{e9}
\xi_f^{(d)}\equiv\mp\exp\left(-i\,2\beta\right)
\frac{\sum\limits_{j=u,c}V_{jq}^\ast\,V_{jb}\,\langle f|{\cal Q}^{jq}|
\overline{B^0_d}\rangle}{\sum\limits_{j=u,c}V_{jq}\,V_{jb}^\ast\,
\langle f|{\cal Q}^{jq}|\overline{B^0_d}\rangle},
\end{equation}
where $2\beta$ is related to the weak $B^0_d$--$\overline{B^0_d}$ mixing
phase and ${\cal Q}^{jq}$ denotes the combination of four-quark operators
and Wilson coefficients appearing in Eq.~(\ref{e4}), 
suffers from large hadronic uncertainties that are 
introduced through the hadronic matrix elements in Eq.~(\ref{e9}).
There is, however, a very important special case where these uncertainties
cancel. It is given if $B_d\to f$ is dominated by a single CKM amplitude.
In that case $\xi_f^{(d)}$ takes the simple form 
\begin{equation}\label{e10}
\xi_f^{(d)}=\mp\exp\left[-i\left(2\beta-\phi_{\mbox{{\scriptsize 
D}}}^{(f)}\right)
\right],
\end{equation}
where $\phi_{\mbox{{\scriptsize D}}}^{(f)}$ is a characteristic decay 
phase that is given by 
\begin{equation}\label{e11}
\phi_{\mbox{{\scriptsize D}}}^{(f)}=\left\{\begin{array}{cc}
-2\gamma&\mbox{for dominant $\bar b\to\bar uu\bar q$ CKM amplitudes 
$(q\in\{d,s\})$ in $B_d\to f$}\\
0&\,\mbox{for dominant $\bar b\to\bar cc\bar q\,$ CKM amplitudes 
$(q\in\{d,s\})$ in $B_d\to f$.}
\end{array}\right.
\end{equation}

Applications and well-known examples of this formalism are the decays
$B_d\to J/\psi\, K_{\mbox{{\scriptsize S}}}$ and $B_d\to\pi^+\pi^-$.
If one goes through the relevant Feynman diagrams contributing to the former
channel one finds that it is dominated to excellent accuracy by the 
$\bar b\to\bar cc\bar s$ CKM amplitude. Therefore the decay phase
vanishes and we have
\begin{equation}\label{e12}
{\cal A}^{\mbox{{\scriptsize mix--ind}}}_{\mbox{{\scriptsize
CP}}}(B_d\to J/\psi\, K_{\mbox{{\scriptsize S}}})=+\sin[-(2\beta-0)].
\end{equation}
Since Eq.~(\ref{e10}) applies to excellent accuracy to the decay
$B_d\to J/\psi\, K_{\mbox{{\scriptsize S}}}$ -- the point is that penguins
enter essentially with the same weak phase as the leading tree
contribution -- it is usually referred to as the ``gold-plated'' mode
to measure the angle $\beta$ of the unitarity triangle\cite{csbs}.

In the case of $B_d\to\pi^+\pi^-$ mixing-induced CP violation would 
measure $-\sin(2\alpha)$ in a clean way through
\begin{equation}\label{e13}
{\cal A}^{\mbox{{\scriptsize mix--ind}}}_{\mbox{{\scriptsize
CP}}}(B_d\to\pi^+\pi^-)=-\sin[-(2\beta+2\gamma)]=-\sin(2\alpha)
\end{equation}
if there were no penguin contributions present. However, such contributions
are there and destroy the {\it clean} relation Eq.~(\ref{e13}). 
The corresponding uncertainties have been estimated recently\cite{kpw} 
to be of the order 25\%. Fortunately, as has been pointed out by
Gronau and London\cite{gl}, the hadronic uncertainties arising from
QCD penguin operators can be eliminated by performing
an isospin analysis involving in addition to $B_d\to\pi^+\pi^-$
also the modes $B_d\to\pi^0\pi^0$ and $B^+\to\pi^+\pi^0$. Following this
approach it is, however, not possible to control also the electroweak
penguin contributions and the related uncertainties in a quantitative
way\cite{dh}. 
As was found some time ago\cite{rfewp} (see also 
Ref.\cite{othersewp}), electroweak penguins may -- in contrast to 
na\"\i ve expectations -- play an important role in certain 
non-leptonic $B$ decays, e.g.\ in $B_s\to\pi^0\phi$. This issue, which 
is due to the fact that the Wilson coefficient of one electroweak penguin 
operator increases strongly with the top-quark mass, led to 
considerable interest in the recent 
literature\cite{othersewp,ghlrewp,PAPI}. Detailed 
analyses\cite{ghlrewp,PAPI} of the $B\to\pi\pi$ isospin 
approach\cite{gl} to determine $\alpha$ show, however, that electroweak
penguins play there only a minor role. The corresponding 
uncertainty $|\Delta\alpha|$ can be estimated\cite{PAPIII} to be smaller 
than $6^\circ$. Consequently electroweak penguins are not expected to 
lead to serious problems in that approach. Since electroweak penguins enter 
in the case of $B_d\to J/\psi\, K_{\mbox{{\scriptsize S}}}$
with the same weak phase as the leading tree contribution they do not 
affect the mixing-induced CP asymmetry measuring $\sin(2\beta)$.

Before we will focus on the $B_s$ system in the next subsection, let me
note one experimental problem of the Gronau--London approach\cite{gl}.
It is related to the fact that it requires a measurement of 
BR$(B_d\to\pi^0\pi^0)$ which is regarded as being very difficult. A recent
analysis by Kramer and Palmer\cite{kp} indicates  
BR$(B_d\to\pi^0\pi^0)\stackrel{<}
{\sim}{\cal O}(10^{-6})$. Therefore it is important to have also alternatives
available to determine $\alpha$ in a clean way. Such alternatives are 
already on the market. For example, Snyder and Quinn\cite{sq} have 
suggested to use $B\to\rho\,\pi$ modes to accomplish this ambitious task. 
Another method\cite{PAPII} has been proposed recently by Buras 
and myself. It requires a simultaneous measurement of 
${\cal A}^{\mbox{{\scriptsize mix--ind}}}_{\mbox{{\scriptsize
CP}}}(B_d\to\pi^+\pi^-)$ and ${\cal A}^{\mbox{{\scriptsize 
mix--ind}}}_{\mbox{{\scriptsize CP}}}(B_d\to K^0\overline{K^0})$
and determines $\alpha$ with the help of the
$SU(3)$ flavor symmetry of strong interactions. Interestingly the 
penguin-induced decay $B_d\to K^0\overline{K^0}$ may exhibit large 
CP-violating asymmetries\cite{rfkokobar} within the Standard Model 
due to interference effects between penguins with internal up- and 
charm-quark exchanges\cite{bf1}. This feature was missed in several 
previous analyses.

Further papers dealing with the penguin uncertainties 
affecting the extraction of $\alpha$ are given in Ref.\cite{alpharef}.
Recently the decays $B_d(t)\to\pi^+\pi^-$, $B^0_d\to\pi^- K^+$,
$B^+\to\pi^+ K^0$ and their CP-conjugates have been considered in 
Ref.\cite{dgr} by using $SU(3)$ flavor symmetry arguments. The corresponding
observables may allow a determination of the angles $\alpha$ and $\gamma$
of the unitarity triangle.

\subsection{The $B_s$ System in Light of $\Delta\Gamma_s$}\label{Bssys}

The situation arising in the $B_s$ system may be quite different from the
$B_d$ case because of the expected sizable width difference\cite{deltagamma}
$\Delta\Gamma_s\equiv\Gamma_H^{(s)}-\Gamma_L^{(s)}$. Here $\Gamma_H^{(s)}$
and $\Gamma_L^{(s)}$ denote the decay widths of the $B_s$ mass eigenstates 
$B_s^{\mbox{{\scriptsize Heavy}}}$ and $B_s^{\mbox{{\scriptsize Light}}}$,
respectively. The major contributions to $\Delta\Gamma_s$, which may be 
as large as ${\cal O}(20\%)$ of the average decay width, originate from 
$\bar b\to \bar cc\bar s$ transitions into final states that are common 
both to $B_s^0$ and $\overline{B_s^0}$. As has been pointed out by 
Dunietz\cite{dunietz}, due to this width difference already 
{\it untagged} $B_s$ rates defined by 
\begin{equation}\label{e14}
\Gamma[f(t)]\equiv\Gamma(B_s^0(t)\to f)+\Gamma(\overline{B^0_s}(t)\to f)
\end{equation}
may provide valuable information about the phase structure of the observable
\begin{equation}\label{e15}
\xi_f^{(s)}=\exp\left(-i\,\Theta_{M_{12}}^{(s)}\right)
\frac{A(\overline{B^0_s}\to f)}{A(B^0_s\to f)},
\end{equation}
where $\Theta_{M_{12}}^{(s)}$ is the weak $B_s^0$--$\overline{B_s^0}$ mixing
phase. This can be seen nicely by writing Eq.~(\ref{e14}) in a more
explicit way as follows:
\begin{equation}\label{e16}
\Gamma[f(t)]\propto\left[\left(1+\left|\xi_f^{(s)}
\right|^2\right)\left(e^{-\Gamma_L^{(s)} t}+e^{-\Gamma_H^{(s)} t}\right)-
2\mbox{\,Re\,}\xi_f^{(s)}\left(e^{-\Gamma_L^{(s)} t}-
e^{-\Gamma_H^{(s)} t}\right)\right].
\end{equation}
In this expression the rapid oscillatory $\Delta M_s\, t$ terms, which
show up in the {\it tagged} rates, cancel\cite{dunietz}. Therefore it 
depends only on the two exponents $e^{-\Gamma_L^{(s)} t}$ and 
$e^{-\Gamma_H^{(s)} t}$, where $\Gamma_L^{(s)}$ and $\Gamma_H^{(s)}$ can 
be determined e.g.\ from the angular distibutions\cite{ddlr} of the decay
$B_s\to J/\psi\,\phi$.

To illustrate these untagged rates in more detail, let me discuss an estimate
of the CKM angle $\gamma$ using {\it untagged} $B_s\to K^+K^-$ and
$B_s\to K^0\overline{K^0}$ decays that has been proposed very recently
by Dunietz and myself\,\cite{fd1}. Using the $SU(2)$ isospin
symmetry of strong interactions to relate the QCD penguin contributions to
these decays (electroweak penguins are color-suppressed in these modes
and thus play a minor role), we obtain
\begin{equation}\label{e17}
\Gamma[K^+K^-(t)]\propto |P'|^2\left[(1-2\,|r|\cos\rho\,\cos\gamma
+|r|^2\cos^2\gamma)e^{-\Gamma_L^{(s)} t}+|r|^2\sin^2\gamma\, 
e^{-\Gamma_H^{(s)} t}
\right]
\end{equation}
and
\begin{equation}\label{e18}
\Gamma[K^0\overline{K^0}(t)]\propto |P'|^2\,e^{-\Gamma_L^{(s)} t},
\end{equation}
where 
\begin{equation}\label{e19}
r\equiv|r|e^{i\rho}=\frac{|T'|}{|P'|}e^{i(\delta_{T'}-\delta_{P'})}.
\end{equation}
Here $P'$ denotes\cite{ghlrsu3} the $\bar b\to\bar s$ QCD penguin amplitude, 
$T'$ is the color-allowed $\bar b\to\bar uu\bar s$ tree amplitude, and 
$\delta_{P'}$ and $\delta_{T'}$ are the corresponding CP-conserving strong 
phases. In order to determine $\gamma$ from the untagged rates 
Eqs.~(\ref{e17}) and (\ref{e18}) we need an additional input that  
is provided by the $SU(3)$ flavor symmetry of strong
interactions. If we neglect the color-suppressed current-current
contributions to $B^+\to\pi^+\pi^0$ we find\cite{ghlrsu3}
\begin{equation}\label{e20}
|T'|\approx\lambda\,\frac{f_K}{f_\pi}\,\sqrt{2}\,|A(B^+\to\pi^+\pi^0)|,
\end{equation}
where $\lambda$ is the Wolfenstein parameter\cite{wolf}, $f_K$ and
$f_\pi$ are the $K$ and $\pi$ meson decay constants, respectively,
and $A(B^+\to\pi^+\pi^0)$ denotes the appropriately normalized 
$B^+\to\pi^+\pi^0$ decay amplitude. Since $|P'|$ is known from
$B_s\to K^0\,\overline{K^0}$, the quantity $|r|=|T'|/|P'|$ can be estimated 
with the help of Eq.~(\ref{e20}) and allows the extraction of $\gamma$
from the part of Eq.~(\ref{e17}) evolving with the exponent 
$e^{-\Gamma_H^{(s)} t}$.

Such an $SU(3)$ flavor symmetry input to determine $\gamma$ is not necessary
in the case of the decays $B_s\to K^{\ast+}K^{\ast-}$ and $B_s\to K^{\ast0}
\overline{K^{\ast0}}$. The angular distributions of these decays
provide much more information than the pseudoscalar-pseudoscalar modes and
therefore the $SU(2)$ isospin symmetry of strong interactions to relate the
QCD penguin contributions suffices to extract\cite{fd1} the CKM angle 
$\gamma$. The time-dependent angular distributions of the decays 
$B_s\to D_s^{\ast+}\,D_s^{\ast-}$, $B_s\to J/\psi\,\phi$ allow the 
determination\cite{fd1} of the Wolfenstein parameter\cite{wolf} $\eta$, and  
time-dependent studies of angular distributions for $B_s$ decays caused
by $\bar b\to\bar cu\bar s$ quark-level transitions extract\cite{fd2}
{\it cleanly} and {\it model-independently} the CKM angle $\gamma$. 
To this end {\it untagged} $B_s$ data samples are sufficient if 
$\Delta\Gamma_s$ is sizable. Even if $\Delta\Gamma_s$ 
should turn out to be too small for such an untagged analysis, 
once $\Delta\Gamma_s\not=0$ has been established experimentally, 
the formulae presented in Refs.\cite{fd1,fd2} have to be used in order 
to determine the CKM phases correctly. 

\subsection{Triangle Relations}\label{trirel}

Let me discuss this class of strategies for extracting CKM phases 
by considering the charged $B$ decays $B^+\to\overline{D^0}K^+$, 
$B^+\to D^0K^+$ and $B^+\to D^0_+K^+$, where $|D^0_+\rangle=
\left(|D^0\rangle+|\overline{D^0}\rangle\right)/\sqrt{2}$
denotes the CP-eigenstate of the neutral $D$ system corresponding to 
eigenvalue $+1$. Using this CP-eigenstate it is an easy exercise to 
derive the amplitude relations
\begin{eqnarray}
\sqrt{2}A(B^+\to D^0_+K^+)&=&A(B^+\to D^0 K^+)+A(B^+\to\overline{D^0}K^+)
\label{e22}\\
\sqrt{2}A(B^-\to D^0_+K^-)&=&A(B^-\to\overline{D^0} K^-)+A(B^-\to D^0K^-)
\label{e23}
\end{eqnarray}
which can be represented as two triangles in the complex plane. Taking into
account moreover the relations
\begin{eqnarray}
A(B^+\to D^0K^+)&=&A(B^-\to\overline{D^0}K^-)\,e^{2i\gamma}\label{e24}\\
A(B^+\to\overline{D^0}K^+)&=&A(B^-\to D^0K^-),\label{e25}
\end{eqnarray}
which arise from the fact that the corresponding transitions 
are pure tree decays, a {\it clean} determination of the 
CKM angle $\gamma$ is possible by using the
triangle relations Eqs.~(\ref{e22}) and (\ref{e23}). This approach has been 
proposed by Gronau and Wyler\cite{gw}, a similar strategy using neutral
$B_d\to DK^\ast$ $(D\in\{D^0,\overline{D^0},D^0_+\})$ 
decays by Dunietz\cite{isi}. Since the corresponding triangles
are expected to be very squashed ones because of certain color-suppression 
effects, and since one is furthermore dealing with the CP-eigenstate 
$|D^0_+\rangle$, this method is unfortunately expected to be very 
challenging from an experimental point of view. 

In some sense the triangle construction discussed above represents a 
proto-type of triangle constructions which have been very popular over
the recent two years. They have been presented in a series of interesting
papers\cite{ghlr} by Gronau, Hern\'andez, London and Rosner who have 
combined the $SU(3)$ flavor symmetry of strong interactions with plausible
dynamical assumptions to derive triangle relations among $B\to\{\pi\pi,
\pi K,K\overline{K}\}$ decay amplitudes. These relations should allow a 
determination of the weak phases of the CKM matrix and of strong final 
state interaction phases by measuring only the corresponding branching
ratios which are all of the order $10^{-5}$. Although this approach is
very attractive at first sight, it has certain limitations. In contrast to
the Gronau--Wyler triangle relations\cite{gw}, the $SU(3)$ triangle relations 
are not valid exactly but suffer from unknown non-factorizable 
$SU(3)$-breaking corrections\cite{ghlrsu3}. Moreover QCD penguins with
internal up- and charm-quark exchanges affect the extraction of the CKM angle
$\beta$ using these relations\cite{bf1}. In addition also electroweak 
penguins\cite{rfewp,othersewp} play an important role in some of these 
constructions. This issue is the subject of the following section. 

\section{The Role of Electroweak Penguins}

While electroweak penguin effects are negligibly small in the CP asymmetry 
of $B_d\to J/\psi\, K_{\mbox{{\scriptsize S}}}$ measuring $\sin(2\beta)$ and 
are expected to be of minor importance in the $SU(2)$ isospin construction 
to determine $\sin(2\alpha)$ from $B\to\pi\pi$ modes as I have noted already 
in Subsection 2.1, electroweak penguins spoil\cite{dh} the 
determination\cite{ghlr} of $\gamma$ using 
$B^+\to\{\pi^0K^+,\pi^+K^0,\pi^+\pi^0\}$ $SU(3)$ triangle relations. 
Several solutions have been proposed in the recent literature to solve this
problem. For example, Gronau et al.\cite{ghlrewp} have proposed an amplitude
quadrangle for $B\to\pi K$ decays that can be used in principle to determine
$\gamma$ irrespectively of the presence of electroweak penguins. Since one
side of this quadrangle is related to BR$(B_s\to\pi^0\eta)={\cal O}(10^{-7})$, 
this approach is very difficult from an experimental point of view. 
Another $SU(3)$-based approach to extract $\gamma$ using the charged $B$ 
decays $B^-\to\{\pi^-\overline{K^0},\pi^0K^-,\eta_8\,K^-\}$ and 
$B^-\to\pi^-\pi^0$, where electroweak penguins are also eliminated, has been 
proposed by Deshpande and He\cite{dh-gamma}. Although this method should
be more promising for experimentalists, it is affected by 
$\eta$--$\eta'$ mixing and other $SU(3)$-breaking effects and  
cannot be regarded as a clean measurement of $\gamma$.

In view of this situation it would be useful to determine the electroweak 
penguin contributions experimentally. The knowledge of the electroweak
penguin amplitudes would allow several predictions, consistency checks 
and tests of certain Standard Model calculations\cite{PAPIII}. 
As Buras and myself have shown in Ref.\cite{PAPI}, if the CKM angle $\gamma$ 
is used as an input, the $\bar b\to\bar s$ electroweak penguin 
amplitude can be determined from the branching ratios for $B^+\to\pi^+K^0$,
$B^+\to\pi^0K^+$, $B^-\to\pi^0K^-$, $B^0_d\to\pi^-K^+$ and 
$\overline{B^0_d}\to\pi^+K^-$ by using only the $SU(2)$ isospin symmetry
of strong interactions. Taking into account that electroweak penguins are
dominated by internal top-quark exchanges and using the $SU(3)$ flavor
symmetry, the $\bar b\to\bar d$ electroweak penguin amplitude can be
determined from the $\bar b\to\bar s$ amplitude\cite{PAPI}. The central
input needed to accomplish this ambitious task -- the CKM angle 
$\gamma$ -- can also be determined approximately with the help of 
the branching ratios for
$B^+\to\pi^+ K^0$, $B^0_d\to\pi^- K^+$, $\overline{B^0_d}\to\pi^+ K^-$
and $B^+\to\pi^+\pi^0$. This approach\cite{PAPIII} should be 
interesting for experiments with limited statistics. In this
respect also the estimate\cite{fd1} of $\gamma$ using untagged $B_s\to
K^+K^-$, $B_s\to K^0\overline{K^0}$ decays discussed in Subsection 2.2 may 
be helpful. Measuring in addition the branching ratios for $B^+\to\pi^0
K^+$ and $B^-\to\pi^0K^-$ the electroweak penguin amplitudes can be 
determined as discussed in Ref.\cite{PAPI}. Once the CKM angle $\gamma$
can be fixed by using absolutely clean methods, e.g.\ the ones
proposed in Refs.\cite{fd2,gw,adk}, the determination of electroweak 
penguins could be improved considerably. That should lead to interesting
and valuable insights into the world of electroweak penguins. An exciting 
future may lie ahead of us!

\section{Acknowledgements}
\par
I would like to thank the organizers of the Workshop for creating such 
a pleasant atmosphere. 

\vspace*{0.5cm}
\par\noindent
{\bf 5.\ References}

\end{document}